\documentclass[fleqn,twoside]{article}
\usepackage{espcrc2}
\usepackage{amssymb}

\usepackage{graphicx}
\usepackage[figuresright]{rotating}

\newcommand{\AmS}{{\protect\the\textfont2
  A\kern-.1667em\lower.5ex\hbox{M}\kern-.125emS}}

\title{Supernova neutrinos}

\author{C. Y. Cardall\address{Physics Division, Oak Ridge National Laboratory, \\
      Oak Ridge, TN 37831-6354, United States of America}%
      \address{Department of Physics and Astronomy, University of Tennessee, \\
	Knoxville, TN 37996-1200, United States of America}%
        \thanks{Oak Ridge National Laboratory is managed by UT-Battelle, LLC, for the DoE under contract DE-AC05-00OR22725.}}

\begin{document}

\begin{abstract}
A nascent neutron star resulting from stellar collapse is a prodigious source of neutrinos of all flavors. While the most basic features of this neutrino emission can be estimated from simple considerations, the detailed simulation of the neutrinos' decoupling from the hot neutron star is not yet computationally tractable in its full glory, being a time-dependent six-dimensional transport problem. Nevertheless, supernova neutrino fluxes are of great interest in connection with the core-collapse supernova explosion mechanism and supernova nucleosynthesis, and as a potential probe of the supernova environment and of some of the neutrino mixing parameters that remain unknown; hence a variety of approximate transport schemes have been used to obtain results with reduced dimensionality. However, none of these approximate schemes have addressed a recent challenge to the conventional wisdom that neutrino flavor mixing cannot impact the explosion mechanism or {\em r-}process nucleosynthesis.
\vspace{1pc}
\end{abstract}

\maketitle

\section{CORE-COLLAPSE SUPERNOVAE}

Stellar collapse results from the evolution of a massive star.
How far a star can burn depends on its total mass $M$; but because of cooling
by neutrino emission from carbon burning onwards, the core of a massive star eventually becomes degenerate, be it composed of O/Ne/Mg ($8\: M_\odot \lesssim M \lesssim 10\: M_\odot$) or Fe ($M \gtrsim 10\:M_\odot$). The core becomes unstable as the ashes of nuclear burning pile up and it approaches the Chandrasekhar mass.
Its inner portion undergoes homologous collapse
(velocity proportional to radius), and the outer portion collapses 
supersonically. 
Electron capture on nuclei is one proximate cause of instability, producing $\nu_e$ that escape freely
until densities 
in the collapsing core
become so high that even neutrinos are 
trapped. 

Collapse halts with a `bounce' soon after the inner core exceeds nuclear 
density, when
a shock wave forms at the boundary between the homologous
and supersonically collapsing regions. The shock begins to move out,
but stalls at some distance beyond the surface of the newly-born neutron star as neutrino emission and endothermic dissociation of heavy 
nuclei sap energy from the shocked material.

Neutrinos dominate the energetics of post-bounce evolution. 
The weakness of their interactions implies that copious neutrino emission is the most efficient means for the hot nascent neutron star to cool. Neutrinos diffuse outward through the dense core on a time scale of seconds and eventually escape with about 99\% of the gravitational energy released during collapse. Some of this is redeposited in neutrino captures behind the stalled shock.
But whether or not this neutrino heating is a proximate cause of shock revival, any definitive account of the explosion mechanism must include careful handling of the neutrinos, for errors in their overwhelming contribution to the energy budget reduce confidence that an energetically subdominant detail like the explosion has been adequately resolved.

What sort of computation is needed to follow the neutrinos' evolution?
 Deep inside the newly-born neutron star, 
 the neutrinos and the fluid are tightly coupled (nearly in equilibrium);  but as 
 neutrinos are transported from inside the neutron star, they go from a nearly isotropic diffusive regime to strongly forward-peaked free-streaming. Heating 
 behind the shock occurs precisely in 
this transition region, and modeling this process accurately ideally involves tracking both the energy and angle dependence of the neutrino distribution functions at every point
in space.

\section{POST-BOUNCE SIMULATIONS}
\label{sec:survey}

A full treatment of this time-dependent six-dimensional neutrino radiation hydrodynamics
problem is a major challenge, and the computational resources necessary to solve it are still just over the horizon.
Nevertheless, much has been 
learned over the years through the simulation of model systems of reduced dimensionality.



Progress on the general trend of increasing total dimensionality (space plus momentum space) handled by simulations in recent years is sketched in Table 1.  It should of course be kept in mind that this does not represent every insight relevant to the explosion mechanism or other aspects of supernova phenomenology, obtained via simulation or otherwise. Nevertheless, the literature bears witness that increases in the dimensionality of sophisticated simulations are intertwined with important advances in the field. (For a previous version of this table that includes entries for some earlier studies, see Ref. \cite{cardall05d}.)

\begin{table*}
\caption{Selected neutrino radiation hydrodynamics milestones in stellar collapse  simulations studying the long-term fate of the shock.
The `Yes' entries in the  `Explosion' column are all marked with an asterisk as a reminder that questions about the simulations---described in the main text---have prevented a consensus about the explosion mechanism. 
`Total dimensions' is the average of `Fluid space dimensions' and  `$\nu$ space dimensions,' added to `$\nu$ momentum space dimensions.'
The abbreviation `N' stands for `Newtonian,' while `PN'---for `Post-Newtonian'---stands for some attempt at inclusion of general relativistic effects, and `GR' denotes full relativity. A space dimensionality in quotes---like `1.5'---denotes an attempt at modeling higher dimensional effects within the context of a lower dimensional simulation. For the fluid, this is a mixing-length prescription in the neutron star (`NS') or the heating region (`HR') behind the stalled shock. For neutrino transport, it indicates one of two approaches: multidimensional diffusion in regions with strong radiation/fluid coupling, matched with a spherically symmetric `light bulb' approximation in weakly coupled regions (`thick/thin'); or the (mostly) independent application of a spherically symmetric formalism/algorithm to separate spatial angle bins (`ray-by-ray').
}
\begin{tabular}{@{}lcccccc}
\hline
Group & Year & Explosion & Total  & Fluid space  & $\nu$ space & $\nu$ momentum
	     \\
	     &       & & dimensions    & dimensions & dimensions & space dimensions \\
\hline
Lawrence & 1989 & Yes$^*$ & `2.25' & `1.5' NS+HR  & 1 & 1   \\
 Livermore \cite{mayle90,mayle91,wilson93} & & & & (GR)  & & (GR) \\
\hline
Lawrence & 1992 & Yes$^*$ & 2 & 2 HR  & 2 & 0  \\
 Livermore \cite{miller93} & & & & (N)  & &(N) \\
\hline
Los  Alamos & 1993  & Yes$^*$ & `1.75' & 2  & `1.5' & 0 \\
\cite{herant94}   & & & & (N)  & thick/thin &(PN) \\
Arizona & 1994 & Yes$^*$ & `1.75' & 2  & `1.5' & 0  \\
\cite{burrows95}  &  & & & (N)  & ray-by-ray & (N) \\
\hline
Florida & 1994  & No & `2.25' & `1.5' NS & 1 & 1 \\
Atlantic \cite{bruenn94,bruenn95} & & & & (GR)  & &(${\cal O}(v/c)$) \\
\hline
Oak Ridge & 1996 & No & `2.5' & 2 & 1 & 1  \\
\cite{mezzacappa98a,mezzacappa98b} & & & & (N)  & &(${\cal O}(v/c)$) \\
\hline
Garching & 2000 & No, Yes$^*$ & 3 & 1 & 1 & 2  \\
\cite{rampp00,rampp02,kitaura03,janka04b,kitaura06} & &  
(8.8 $M_\odot$)& & (N)  & &(${\cal O}(v/c)$) \\
Oak Ridge & 2000 & No & 3 & 1 & 1 & 2  \\
\cite{mezzacappa93b,mezzacappa93c,mezzacappa99,liebendoerfer00,mezzacappa01} & & & & (N)  & &(${\cal O}(v/c)$) \\
Arizona & 2002 & No & 3 & 1 & 1 & 2  \\
\cite{burrows00,thompson03} & & & & (N)  & &(${\cal O}(v/c)$) \\
\hline
Oak Ridge & 2000  & No & 3 & 1 & 1 & 2 \\
\cite{mezzacappa93b,mezzacappa93c,mezzacappa99,liebendoerfer01a,liebendoerfer02,liebendoerfer04} & & & & (GR)  & &(GR) \\
\hline
Los  Alamos & 2002 & Yes$^*$ & `2.5' & 3  & `2' & 0  \\
\cite{herant94,fryer02}  &  & & & (N)  & thick/thin &(PN) \\
\hline
Garching & 2002 & No, Yes$^*$ & `3.75' & 2 & `1.5' & 2 (${\cal O}(v/c)$,  \\
\cite{rampp02,janka04b,buras03,janka02,janka04a,buras06,buras06b} & & (11.2 $M_\odot$)
& & (PN)  & ray-by-ray & PN) \\
\hline
Arizona & 2005 & Yes$^*$ & 3 & 2 & 2 & 1 (${\cal O}(1)$,  \\
\cite{walder04,burrows06,burrows06b} & & 
& & (N)  &  & N) \\
\hline
Florida & 2006  & Yes$^*$ & `2.75' & 2 & `1.5' & 1 \\
Atlantic \cite{bruenn06} & & & & (N)  & &(${\cal O}(v/c)$) \\
\hline
\end{tabular}\\[2pt]
\label{history}
\end{table*}

That spherical symmetry would be an incomplete description was suggested both by observations of supernova SN1987A and by profiles of fluid variables obtained from spherically symmetric simulations. Working mostly with mock-ups of multidimensional instabilities within spherical symmetry, one group suggested that a `doubly-diffusive instability' within the neutron star---which could boost neutrino luminosities---was more important (even crucial) to an explosion than convection immediately behind the shock, which enhances the efficiency of neutrino heating \cite{mayle90,mayle91,wilson93,miller93}. Other studies, sacrificing the energy dependence of neutrino distributions in exchange for genuine multidimensional fluid dynamics, found the effects of post-shock convection to be more important \cite{burrows95} and even robust \cite{herant94,fryer02}. Subsequent work could not confirm the presence of doubly-diffusive instabilities \cite{bruenn95,bruenn96,bruenn04,dessart06}, and the operation of more conventional convection within the neutron star appears to be either suppressed by neutrino transport \cite{mezzacappa98a} or unimportant to explosion dynamics \cite{buras03,dessart06}. And echoing earlier pessimism \cite{miller93}, later work retaining neutrino energy dependence also deflated enthusiasm for the capacity of post-shock convection to robustly bring about neutrino-heated explosions \cite{mezzacappa98b,buras03}.

The nagging qualitative difference between spatially multidimensional
simulations with different neutrino transport approximations 
motivated interest in the possible importance of more complete neutrino
transport: might the retention of both the energy
{\em and} angle dependence of the neutrino distributions improve the chances of explosion, as preliminary `snapshot' studies suggested \cite{messer98,burrows00}?
Of necessity, the first such simulations were performed in
spherical symmetry, which nevertheless represented an advance to a total dimensionality of 3 (see Table 1).
Results from
three different groups are in accord: Spherically symmetric models of iron core collapse 
do not explode, even with solid neutrino transport 
\cite{rampp00,mezzacappa01,thompson03} and general relativity \cite{liebendoerfer01a,liebendoerfer04}. Recently, however, it has been shown that the more modest O/Ne/Mg cores of the lightest stars to undergo core collapse (8-10 $M_\odot$) may explode in spherical symmetry \cite{kitaura03,janka04b,kitaura06}.

In an extension to two space dimensions, one of these groups deployed their spherically symmetric energy- and angle-dependent neutrino transport capability \cite{rampp02} along separate radial rays, with partial coupling between rays \cite{janka02,buras06}. 
Initial results---from axisymmetric simulations with a restricted angular domain---were negative with regards to explosions (in spite of the salutary effects of convection, and also rotation) \cite{buras03,buras06}, apparently supporting the results of Ref. \cite{mezzacappa98b}. 
An explosion was seen in one simulation \cite{janka02,janka03,buras06}
in which
certain terms in the neutrino transport equation corresponding to Doppler shifts
and angular aberration due to fluid motion were dropped; this simulation also yielded a neutron star mass and nucleosynthetic consequences in better agreement with observations than the exploding simulations of the 1990s \cite{herant94,burrows95}, arguably because of more accurate neutrino transport in the case of both observables. 
The continuing lesson
is that getting the details of the neutrino transport right makes a difference.

In addition to accurate neutrino transport, low-mode ($\ell = 1,2$) instabilities that can develop only in simulations allowing the full range of polar angles may ultimately make decisive differences. 

One example is an explosion, obtained with the `ray-by-ray' code described above \cite{janka04a,janka04b,buras06b}, of one of the lowest mass stars (11.2 $M_\odot$) to have an iron core: an explosion was seen when the full 180$^{\mathrm{o}}$ in polar angle was used, but not when the grid covered only 90$^{\mathrm{o}}$. Explosions of 11 $M_\odot$ and 15 $M_\odot$ progenitors have also been obtained by another group on a 180$^{\mathrm{o}}$ grid using another ray-by-ray code that maintains only the neutrino energy dependence (flux-limited diffusion) \cite{bruenn06}.\footnote{In addition to the shock instability to be mentioned shortly, these authors present evidence that the use of a nuclear network rather than a representative heavy nucleus contributes materially to a successful explosion.} These achievements were presaged by an earlier study that demonstrated the tendency for convective cells to merge to the lowest order allowed by the spatial domain \cite{herant92}, and especially by the more recent discovery of a new standing accretion shock instability \cite{blondin03}. It is not yet entirely clear whether the standing accretion shock instability simply results from a standing acoustic wave in the cavity between the neutron star and the shock \cite{blondin06}, or is instead an advective/acoustic cycle in the same region \cite{foglizzo02,ohnishi06,foglizzo06b}; but in any case a suspicion is becoming widely established (e.g. \cite{scheck04,burrows06,buras06b,bruenn06}) that a shock instability is operative in the full complexity of the supernova environment that generates asymmetry independently of convection \cite{blondin03,foglizzo06,yamasaki06}. 
These shock-instability-induced global asymmetries seem to be sufficient to account for observed asphericities that have often been attributed to rotation and/or magnetic fields \cite{janka05,burrows06b}. 

In another example, this shock instability is acknowledged but claimed to only set the stage for another low-mode instability that is more directly related to the explosion, namely, $\ell = 1$ (and, later and subdominantly, $\ell = 2$) $g-$mode oscillations of the nascent neutron star \cite{burrows06,burrows06b}. Excited by accretion streams funneled onto the neutron star by the flow structures resulting from the shock instability, the claim is that these oscillations of the core radiate acoustic waves that steepen into shocks which then deposit sufficient energy at larger radii to drive the explosion. This effect has so far only been seen by one group, with a code whose neutrino transport is two-dimensional in space but that retains only the energy dependence of the neutrino distributions (flux-limited diffusion).\footnote{The neutrino transport also does not account for most velocity-dependent terms or inelastic neutrino scattering.} It would not have been seen in simulations that either exploded by the neutrino-driven mechanism before this acoustic mechanism had time to develop, or that were stopped `prematurely' when it became clear that a neutrino-driven explosion was not forthcoming. On the other hand, at least one other group \cite{bruenn06} has now run simulations out to the requisite times with a ray-by-ray flux-limited diffusion code, and the acoustic mechanism has not been observed. It might be claimed that this latter code's use of spherical coordinates artificially pins the core to the origin and prevents the necessary oscillations \cite{burrows06b}. However, an order-of-magnitude mismatch between the frequencies of the $g-$mode core oscillations and the shock instability that would drive it is a bit puzzling. Confirmation by other codes will be needed to eliminate the possibility of the acoustic mechanism being merely an unphysical simulation artifact. 

Surely every `Yes' entry in the explosion column of Table 1 has been hailed in its time as `the answer' (at least by some!), and as a community we cannot help hoping once again that these recent developments mark the turning of a corner; but important work remains to verify if this is the case. At least two groups are pursuing neutrino transport in two and three space dimensions that retain the full energy and angle dependence of the neutrino phase space \cite{cardall04,cardall05b,cardall05d,livne04,hubeny06}.

\section{SUPERNOVA NEUTRINO SIGNALS}

Whether it be by playing the spoiler, acting as agent of explosion, or setting the conditions necessary to the operation of some other mechanism, neutrinos have direct or indirect roles in the many explosion scenarios discussed over the years---and detailed simulation of collapse and the second or so after bounce will allow a neutrino signal detected from a Galactic supernova to serve as an unrivaled diagnostic tool.

\begin{figure}
\includegraphics[width=3in]{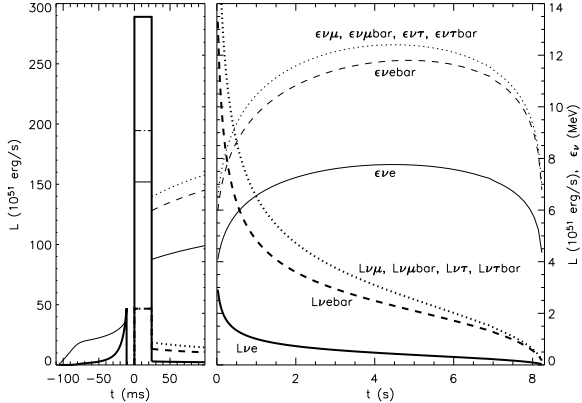}
\vspace{-40pt}
\caption{\small Crudely estimated neutrino luminosities (thick) and characteristic energies (thin). Note the change in time and luminosity (but not energy) scales between the left and right panels. The left panel is a close-up of infall and bounce at $t=0$.}
\label{neutrinos}
\vspace{-15pt}
\end{figure}

But apart from its potential dynamic and diagnostic roles in connection with the explosion,  supernova neutrino emission is of intrinsic interest as an energetically dominant feature of stellar collapse. 
Figure \ref{neutrinos}, obtained from a simple and self-contained model \cite{cardall07}, serves as a recognizable if very rough caricature of neutrino light curves produced by detailed models \cite{buras06,liebendoerfer04,thompson03,prakash01a}. Apparent are a gradual increase in $\nu_e$ luminosity during infall, which then drops with neutrino trapping; a burst dominated by $\nu_e$ just after core bounce; and a cooling phase characterized by emission of $\nu$ and $\bar\nu$ of all flavors over many seconds. The lower luminosities and average energies of $\nu_e$ and $\bar\nu_e$ during the cooling phase result from their participation in charged-current interactions unavailable to the other species at these energies, with $\nu_e$ differing from $\bar\nu_e$ because $\nu_e$-capturing neutrons are more abundant than $\bar\nu_e$-capturing protons.

At present, detailed simulations focusing on the explosion assume massless neutrinos, relying on the standard expectation that the effective mass from neutrino forward scattering off electrons suppresses flavor mixing in the high-density region where the explosion is launched. Before reaching terrestrial detectors, however, neutrinos will pass through flavor-changing MSW resonances at lower densities in the stellar envelope, and possibly also experience flavor-changing effects while traversing the earth. Hence collapse simulations provide neutrino spectra as input to studies that take these effects into account in determining possible signals in terrestrial neutrino detectors. To the extent that generic energy- and time-dependent features of the signals can be motivated by the large-scale simulations, such studies provide a possible basis for learning---from neutrino detection of a future Galactic supernova---about remaining unknowns surrounding neutrino flavor mixing. For instance, observation of the $\nu_e$ burst could rule out $\sin^2\theta_{13} \gtrsim 10^{-3}$ and a `normal' mass hierarchy \cite{kachelriess05}. If $\sin^2\theta_{13} \gtrsim 10^{-3}$ it may also be possible to distinguish between `normal' and `inverted' mass hierarchies on the basis of whether it is the $\nu_e$ or $\bar\nu_e$ signal that is affected by the passage of the supernova shock wave through the $\Delta m^2_{13}$ resonance in the supernova envelope \cite{schirato02,fogli03}, or by the neutrinos' traversal through the earth \cite{dighe03}. Passage through the earth could also affect both the $\nu_e$ and $\bar\nu_e$ channels even if $\sin^2\theta_{13} \lesssim 10^{-5}$ \cite{dighe03}.

\section{NEW EFFECTS AT SMALL $\Delta m^2$?}

Again, the conventional wisdom is that the MSW effect suppresses flavor mixing until the  resonances corresponding to the solar and atmospheric neutrino mass-squared differences are reached far out in the supernova envelope; however, this standard expectation may in fact be wrong. If the off-diagonal contributions of neutrino-neutrino forward scattering dominate, both neutrinos {\em and} antineutrinos can mix maximally over a significant range of neutrino energy---and this could occur deep in the supernova environment, even with the known `small' mass-squared differences \cite{fuller06}. In fact a couple of types of collective neutrino mixing behavior of potential relevance to supernovae have been identified in both analytic and numerical studies \cite{duan06,duan06b,hannestad06,raffelt07}, though their ultimate impact is not yet clear. It may well be that, as a result of nonlinear effects associated with neutrino-neutrino forward scattering, flavor mixing will have to be piled on top of the challenge of high dimensionality that supernova modelers already face in simulating neutrino transport.

\def\aap{Astron. Astrophys. }
\def\aasma{Am. Astron. Soc. Meet. Abs. }
\def\apj{Astrophys. J. }
\def\apjl{Astrophys. J. Lett. }
\def\apjs{Astrophys. J. Supp. Ser. }
\def\araa{Annu. Rev. Astron. Astrophys. }
\def\baas{Bull. Am. Astron. Soc. }
\def\jcam{J. Comp. Appl. Math. }
\def\nat{Nature }
\def\npa{Nucl. Phys. A }
\def\npbproc{Nucl. Phys. B, Proc. Suppl. }
\def\pr{Phys. Rep. }
\def\prd{Phys. Rev. D }
\def\prl{Phys. Rev. Lett. }
\def\prv{Phys. Rev. }


\end{document}